\documentclass[manuscript]{emulateapj}

\usepackage{amsmath}
\usepackage{color}

\newcommand{\acp}{Atmos. Chem. Phys.}

\newcommand{\chemrev}{Chem. Rev.}
\newcommand{\cp}{Chem. Phys.}
\newcommand{\jms}{J. Mol. Spectr.}

\newcommand{\pccp}{Phys. Chem. Chem. Phys.}

\newcommand{\jpcrd}{J. Phys. Chem. Ref. Data}
\newcommand{\molphys}{Mol. Phys.}
\newcommand{\jpca}{J. Phys. Chem. A}
\newcommand{\bbpc}{Ber. Bunsen. Phys. Chem.}
\newcommand{\rsi}{Rev. Sci. Instrum.}
\newcommand{\natchem}{Nature Chem.}
\newcommand{\arpcc}{Annu. Rep. Prog. Chem. C}
\newcommand{\jcsft}{J. Chem. Soc. Faraday Trans.}
\newcommand{\science}{Science}

\shorttitle{Reactivity of OH and CH$_3$OH between 22 and 64 K}
\shortauthors{Anti\~nolo et al.}

\begin{document} 

\title{Reactivity of OH and CH$_3$OH between 22 and 64 K:\\ Modelling the gas phase production of CH$_3$O in Barnard 1b}

\author{M.~Anti\~nolo$^{1,2}$, M.~Ag\'undez$^3$, E.~Jim\'enez$^{1,2}$, B. Ballesteros$^{1,2}$, A. Canosa$^{4}$, G. El Dib$^4$, J. Albaladejo$^{1,2}$, and J.~Cernicharo$^3$} 

\affil{
$^1$ Departamento de Qu\'imica F\'isica, Facultad de Ciencias y Tecnolog\'ias Qu\'imicas, Universidad de Castilla-La Mancha\\ Avda. Camilo Jos\'e Cela s/n, E-13071, Ciudad Real, Spain; \email{elena.jimenez@uclm.es} \\
$^2$ Instituto de Combusti\'on y Contaminaci\'on, Universidad de Castilla-La Mancha. Camino de Moledores s/n, E-13071, Ciudad Real, Spain \\
$^3$ Instituto de Ciencia de Materiales de Madrid, CSIC. C/ Sor Juana In\'es de la Cruz 3, E-28049 Cantoblanco, Spain \\
$^4$ Institut de Physique de Rennes, UMR 6251 CNRS-Universit\'e de Rennes 1\\ Campus de Beaulieu, B$\hat{\rm a}$t 11C, 263 Av. G\'en\'eral Leclerc, F-35042, Rennes, France \\
}

\email{elena.jimenez@uclm.es; andre.canosa@univ-rennes1.fr}

\begin{abstract}

In the last years, ultra-low temperature chemical kinetic experiments have demonstrated that some gas-phase reactions are much faster than previously thought. One example is the reaction between OH and CH$_3$OH, which has been recently found to be accelerated at low temperatures yielding CH$_3$O as main product. This finding opened the question of whether the CH$_3$O observed in the dense core Barnard 1b could be formed by the gas-phase reaction of CH$_3$OH and OH. Several chemical models including this reaction and grain-surface processes have been developed to explain the observed abundance of CH$_3$O with little success. Here we report for the first time rate coefficients for the gas-phase reaction of OH and CH$_3$OH down to a temperature of 22 K, very close to those in cold interstellar clouds. Two independent experimental set-ups based on the supersonic gas expansion technique coupled to the pulsed laser photolysis-laser induced fluorescence technique were used to determine rate coefficients in the temperature range 22-64 K. The temperature dependence obtained in this work can be expressed as $k$(22-64 K) = $(3.6\pm0.1)\times10^{-12}(T/ 300~{\rm K})^{-(1.0\pm0.2)}$ cm$^3$ molecule$^{-1}$ s$^{-1}$. Implementing this expression in a chemical model of a cold dense cloud results in CH$_3$O/CH$_3$OH abundance ratios similar or slightly lower than the value of $\sim3\times10^{-3}$ observed in Barnard 1b. This finding confirms that the gas-phase reaction between OH and CH$_3$OH is an important contributor to the formation of interstellar CH$_3$O. The role of grain-surface processes in the formation of CH$_3$O, although it cannot be fully neglected, remains controversial.

\end{abstract}


\section{Introduction} \label{sec:intro}

Cold interstellar clouds are known to contain a rich variety of molecules whose synthesis has long been thought to rely on gas-phase ion-neutral chemical reactions \citep{her1989}. Since the 1990s, chemical kinetic experiments have demonstrated that many gas-phase neutral-neutral reactions in which at least one of the reactants is a radical are very rapid at low temperatures \citep{can2008,smi2013} and thus play an important role in interstellar chemistry \citep{smi2004}. Because cold interstellar clouds have extremely low temperatures, usually around 10 K, only gas-phase chemical reactions which are exothermic and barrierless are considered in reaction networks currently used to model the chemistry of cold interstellar clouds \citep{agu2013}.

Recently, the reaction of hydroxyl (OH) radicals and methanol (CH$_3$OH) has been experimentally found to be very rapid at temperatures down to 56 K, despite the presence of an energy barrier, which is probably surpassed by quantum tunnelling \citep{sha2013,gom2014}. This finding has important potential implications for the chemistry of the interstellar medium. Indeed, the empirical evidence of a chemical reaction that overcomes an energy barrier at low temperatures opens the possibility to find other reactions with a similar behaviour (see, e.g., \citealt{sim2013}) that could be potentially important for interstellar chemistry albeit to date they are not included in chemical models. Hitherto, reaction kinetic databases used to model interstellar chemistry, such as UMIST and KIDA \citep{mce2013,wak2015}, are largely based on data obtained at temperatures above 200 K \citep{atk2006}. In particular, the rate constant of the reaction between OH and CH$_3$OH, which can be represented by the expression $2.85\times10^{-12} \exp{(-345/T)}$ cm$^3$ molecule$^{-1}$ s$^{-1}$ in the temperature range 210-300 K \citep{atk2006}, was previously presumed to be negligible at interstellar temperatures.

There are two possible pathways for the titled reaction:
\begin{subequations} \label{reac:1}
\begin{align}
\rm CH_3OH + OH & \rm \rightarrow CH_3O + H_2O, \label{reac:1a} \\
\rm CH_3OH + OH & \rm \rightarrow CH_2OH + H_2O. \label{reac:1b}
\end{align}
\end{subequations}
The branching ratios of channels (\ref{reac:1a}) and (\ref{reac:1b}) were reported to depend on temperature between 70 and 900 K \citep{sha2013}. At temperatures above 250 K, the formation of hydroxymethyl (CH$_2$OH) radicals is favoured, while at much lower temperatures the formation of CH$_3$O via reaction (\ref{reac:1a}) was reported to be the major reaction pathway \citep{sha2013}. At 82 K, it was found experimentally that the formation of CH$_3$O in the reaction of OH with methanol occurs at a similar rate than the removal of OH radicals \citep{sha2013}. Master equation calculations performed by \citet{sha2013} corroborated the experimental observations and indicated that at 70 K, CH$_3$O radicals are expected to be formed with a $>99\%$ yield. For that reason, channel (\ref{reac:1a}) and its rate coefficient, $k_{1a}$, are referred henceforth to as reaction~(\ref{reac:1}) and $k_1$, respectively. \citet{cer2012} recently detected CH$_3$O toward the cold dense cloud Barnard 1b (B1-b) where the gas kinetic temperature is 10-15 K. Taking into account that methanol is $\sim300$ times more abundant than CH$_3$O in B1-b \citep{obe2010,cer2012}, Cernicharo et al. suggested that the gas-phase reaction between OH and CH$_3$OH emerges as a potential efficient way to form CH$_3$O in this source.

\begin{deluxetable}{lllcrl}
\tablecaption{Summary of the experimental conditions in the employed CRESU systems \label{tab:1}} \tablewidth{0pc}
\startdata \hline \hline
\multicolumn{1}{l}{Bath} & \multicolumn{1}{c}{$P_{res}$} & \multicolumn{1}{c}{$P_{cham}$} & \multicolumn{1}{c}{$M$} & \multicolumn{1}{c}{$T$} & \multicolumn{1}{l}{CRESU} \\
\multicolumn{1}{l}{gas} & \multicolumn{1}{c}{(mbar)} & \multicolumn{1}{c}{(mbar)} & \multicolumn{1}{c}{} & \multicolumn{1}{c}{(K)} & \multicolumn{1}{l}{system} \\
\hline
He              & 337.0 & 0.621 & 6.1$\pm$0.2    & 22.4$\pm$1.4 & UCLM\tablenotemark{a} \\
N$_2$/He  & 127.1 & 0.296 & 5.0$\pm$0.1    & 42.5$\pm$1.3 & UCLM\tablenotemark{b} \\
N$_2$       & 107.3 & 0.180 & 5.09$\pm$0.04 & 47.7$\pm$0.6 & UR1\tablenotemark{c} \\
N$_2$       & 136.2 & 0.279 & 4.9$\pm$0.1     & 51.6$\pm$1.7 & UCLM\tablenotemark{b} \\
Ar              & 28.20 & 0.390 & 3.73$\pm$0.04 & 52.2$\pm$0.9 & UR1\tablenotemark{c} \\
N$_2$       & 41.75 & 0.184 & 4.36$\pm$0.05 & 61.0$\pm$1.0 & UR1\tablenotemark{c} \\
N$_2$       & 41.67 & 0.183 & 4.2$\pm$0.1     & 64.2$\pm$1.7 & UCLM\tablenotemark{b}
\enddata
\tablecomments{$P_{res}$ is the pressure in the reservoir, $P_{cham}$ the pressure in the chamber, $M$ the Mach number, and $T$ the temperature of the jet.}
\tablenotetext{a}{Pulsed mode (see more details in \citealt{jim2015a}).} 
\tablenotetext{b}{Continuous mode (see \citealt{jim2015b}).}
\tablenotetext{c}{\citet{sim1994}; \citet{can2008}.}
\end{deluxetable}

Soon after reaction~(\ref{reac:1}) was empirically found to be accelerated at low temperatures with respect to room temperature \citep{sha2013}, the chemistry of cold dark clouds has been revisited from a theoretical point of view \citep{vas2013,reb2014,rua2015,bal2015,ach2015,kal2015}. These studies were mainly dedicated to explore the formation of complex organic molecules, such as CH$_3$OCH$_3$ and HCOOCH$_3$, although some of them discussed also the formation of CH$_3$O \citep{vas2013,rua2015,bal2015,ach2015}. The rate coefficient $k_1$ adopted by these authors was in consonance with the values reported by \citet{sha2013}. The value of $k_1$ measured by \citet{sha2013} at 63 K was ca. $4\times10^{-11}$ cm$^3$ molecule$^{-1}$ s$^{-1}$ and calculations performed by these authors indicate that by 20 K the rate coefficient has reached the collision limit ($3\times10^{-10}$ cm$^3$ molecule$^{-1}$ s$^{-1}$). Extending the temperature range down to 56 K, \citet{gom2014} measured a rate coefficient ca. $5\times10^{-11}$ cm$^3$ molecule$^{-1}$ s$^{-1}$ and, using their kinetic data in the range 56-88 K, reported an extrapolated value of $k_1$ at 0 K of ca. $6\times10^{-11}$ cm$^3$ molecule$^{-1}$ s$^{-1}$, which suggests a less pronounced increase of $k_1$ with decreasing temperature than predicted by \citet{sha2013}. The value of $k_1$ adopted by chemical models of cold dark clouds that have addressed the formation of CH$_3$O has been either $4\times10^{-11}$ cm$^3$ molecule$^{-1}$ s$^{-1}$ or $3\times10^{-10}$ cm$^3$ molecule$^{-1}$ s$^{-1}$.

In this paper, we present the first determination of rate coefficients for the reaction of OH with CH$_3$OH at temperatures down to 22 K, which allows having a first complete picture of the temperature dependence of $k_1$ at temperatures below 50 K and much closer to the typical temperatures, $\sim10$ K, found in dark molecular clouds. In the light of the new rate coefficients reported here, we also re-evaluate whether the gas-phase route involving reaction~(\ref{reac:1a}) is by itself able to explain the CH$_3$O abundance observed in B1-b. For that purpose, we model the chemistry of cold dense clouds adopting the extrapolated value of $k_1$ at 10 K from the temperature dependence observed in this work.

\section{Methods} \label{sec:methods}

\subsection{Measurement of the rate coefficients for reaction~(\ref{reac:1}) at ultra-low temperatures}

\subsubsection{Ultra-cold environment:\\ Uniform supersonic gas expansion technique}

The CRESU (\emph{Cin\' etique de R\' eaction en Ecoulement Supersonique Uniforme}, which means Reaction Kinetics in a Uniform Supersonic Flow) technique has been used in this work to cover the temperature range 22-64 K. Two independent apparatus were employed. These are the CRESU apparatus recently constructed in the University of Castilla-La Mancha (hereafter referred as UCLM system) and the CRESU system available in the University of Rennes 1 (hereafter referred as UR1 system). The UCLM CRESU was used to investigate the kinetics of OH and methanol at four temperatures (22 K, 42 K, 52 K and 64 K), whereas the UR1 CRESU system operated at three temperatures (48 K, 52 K and 61 K).

As both systems have been previously described elsewhere \citep{jim2015a,jim2015b,sim1994,can2008} only some remarks are mentioned here. In both CRESU systems, a stainless steel chamber is connected to a pumping system to generate low pressures inside ($P_{cham}$). A supersonic flow is achieved by isentropic expansion of a buffer gas through a specifically designed Laval nozzle separating a movable stagnation reservoir maintained at room temperature from the main vacuum chamber. The quality and the physical conditions of the flow directly depend on the geometry of the Laval nozzle on the one hand and on the nature and the flow rate of the buffer gas on the other hand. The gas flows through the reservoir at a constant flow rate, providing a constant background pressure $P_{cham}$, when using the suitable pumping speed. The UCLM system can operate with a gas flow in either pulsed or continuous mode, while the UR1 system is operating exclusively in the latter one. In the UCLM pulsed mode ($T=22$ K), the movable expansion system also included a rotary disk (aerodynamic chopper) in order to pulse the gas inside the Laval nozzle, while in the continuous mode ($T=42, 52, 64$ K) the gas is constantly flowing by one of the apertures of the aerodynamic chopper maintained at rest. In the UR1 system, there is no rotary disk in the movable expansion system. In Table~\ref{tab:1} a summary of the experimental conditions is presented for both CRESU systems. In the UCLM system, the same convergent-divergent Laval nozzle was operated under different physical conditions to achieve the desired temperatures and gas densities. For temperatures higher than 22 K, the bath gas (helium) was changed by (nitrogen) or a mixture of both gases, as previously described in \citet{jim2015b}. In the UR1 system, nitrogen or argon was used as a bath gas to obtain the three different temperature conditions, each of them achieved by using a different Laval nozzle.

\subsubsection{Kinetic technique and\\ determination of the rate coefficients $k_1$}

For carrying out the kinetic experiment of reaction~(\ref{reac:1}), firstly the OH radicals have to be generated in situ. In this work, the Pulsed Laser Photolysis (PLP) of gaseous H$_2$O$_2$,
\begin{equation}
\rm H_2O_2 + {\emph h\nu} \rightarrow 2~OH, \label{reac:2}
\end{equation}
was the source of OH radicals in both CRESU systems.

The photolysis wavelengths were 248 nm (from a KrF excimer laser in UCLM) and 266 nm (from the fourth harmonic of a Nd-YAG laser in UR1). Gaseous H$_2$O$_2$ was introduced into the reservoir by passing a small flow of the bath gas through a concentrated solution of H$_2$O$_2$ contained in a glass bubbler. Aqueous solution of H$_2$O$_2$, which was provided by Sharlab at a concentration of 50 \% v/v, was further purified by bubbling helium or nitrogen through the liquid for a few days in order to remove water at least partly. Concentrations of about 70 \% were obtained by this method for H$_2$O$_2$ in the solution. Liquid samples of methanol (with a purity $\geq99.8\%$ in UCLM and HPLC grade in UR1) were degassed prior to \textbf{its} use. The introduction of methanol in the reservoir was made differently in UCLM and in UR1. In the UCLM system, mixtures of methanol vapour and the bath gas were prepared and stored in two 20 L glass bulbs prior to the experiments. Then, the content of the storage bulbs was directed into the reservoir by means of a calibrated mass flow controller. In the UR1 set-up, gaseous methanol was introduced in the reservoir similarly to H$_2$O$_2$, i.e., by passing a bath gas identical to the main buffer gas through a bubbler containing liquid CH$_3$OH. The temperature and pressure were measured in the bubbler during experiments. The partial flow of methanol was then obtained knowing the bath gas flow rate as well as the vapour pressure of CH$_3$OH at the bubbler temperature (see \citealt{eld2013} for details). In both CRESU systems, a third flow of the buffer gas was introduced into the reservoir by a separate entry port to reach the desired reservoir pressure $P_{res}$. This flow is the main contributor to the total gas flow through the reservoir.

\begin{figure}
\centering
\includegraphics[angle=0,width=\columnwidth]{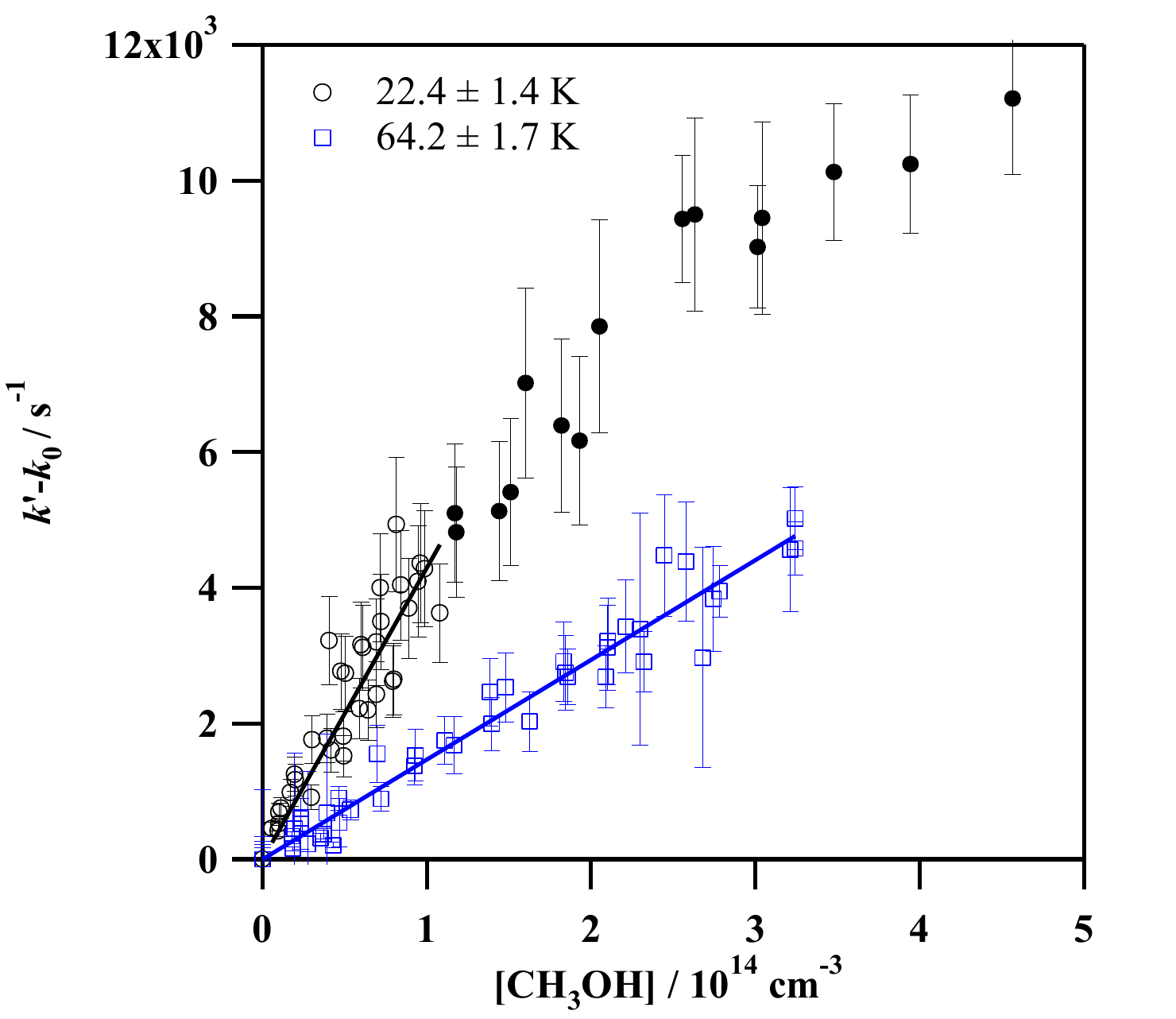} \caption{Plot of $k' - k_0$ versus [CH$_3$OH] at 22 and 64 K. Full black circles, which correspond to kinetic data at 22 K that are influenced by methanol clustering ([CH$_3$OH] $>1\times10^{14}$ molecule cm$^{-3}$, see text), have not been considered in the fit of Equation~(\ref{eq:5}).} \label{fig:1}
\end{figure}

In the presence of an excess of methanol, i.e., under pseudo-first order conditions where [CH$_3$OH] $\gg$ [OH] and [H$_2$O$_2$] $\gg$ [OH], the reaction scheme is described by reaction~(\ref{reac:1}), the reaction of OH with H$_2$O$_2$ (reaction~\ref{reac:3}), and other losses of OH, such as diffusion or reaction with impurities, if any (reaction~\ref{reac:4}): 
\begin{align}
\rm CH_3OH + OH & \rm \rightarrow products, \tag{\ref{reac:1}} \\
\rm H_2O_2  + OH & \rm \rightarrow H_2O + HO_2, \label{reac:3} \\
\rm OH                   & \rm \rightarrow other~losses. \label{reac:4}
\end{align}
Under our experimental conditions, reaction~(\ref{reac:4}) can be neglected since in the absence of methanol the loss of OH radicals is governed by reaction with H$_2$O$_2$, which is in large excess.

To obtain the kinetic information, the OH radicals formed in reaction~(\ref{reac:2}) are excited at 282 nm (radiation from a frequency-doubled dye laser pumped either by a XeCl excimer laser in UCLM or a 532 nm Nd-YAG laser in UR1). The laser induced fluorescence (LIF) from excited OH radicals was detected at ca. 310 nm by a photomultiplier tube as a function of reaction time. The timescale of the reaction, which depends on the rate coefficient $k_1$ and the concentration of methanol introduced into the reservoir, must be short enough to be smaller or of the same order of magnitude than the hydrodynamic time during which the supersonic flow is kept uniform (usually several hundreds of $\mu$s, the exact value being specific to each flow condition validated for a given Laval nozzle). At short reaction times (tens of $\mu$s), the LIF signal from excited OH increases due to rotational relaxation, so in the analysis of the OH temporal profiles, only the decay of the LIF signal at $t>20~\mu$s is considered. The analysis of the exponential decays yields the pseudo-first order rate coefficients, $k'$, which is linearly related to [CH$_3$OH] under pseudo-first order conditions:
\begin{equation}
k' = k_0 + k_1(T) \rm [CH_3OH], \label{eq:5}
\end{equation}
where $k_1(T)$ is the bimolecular rate coefficient for reaction~(\ref{reac:1}) at a given temperature and $k_0$ is the measured rate coefficient in the absence of methanol at the same temperature, when only reactions~(\ref{reac:3}) and (\ref{reac:4}) take place. $k_0$ ranged between 2800 and 8400 s$^{‑1}$. This meant a contribution of $k_0$ to $k'$ very variable, between 40 and 95~\%, although it typically did not exceed 70~\%. Varying the concentration of methanol and maintaining the concentration of the precursor of OH radicals constant, $k_1(T)$ can be obtained from the slope of the plot of $k'$ (or $k'-k_0$) versus [CH$_3$OH]. In the present study, we have chosen the second option, because it is the only way to compare results obtained at the same temperature in independent experiments, in which $k_0$ may differ up to 2000 s$^{-1}$. Methanol concentrations were varied in UR1 by changing the pressure inside the CH$_3$OH bubbling column using a micrometric valve. In UCLM, [CH$_3$OH] was varied by changing the mass flow rate of the diluted mixture from the storage bulb. Examples of plots of $k'-k_0$ versus [CH$_3$OH] are presented in Fig.~\ref{fig:1}. In the UCLM system, up to 48 kinetic experiments were carried out at a single temperature, using different mixing ratios of the CH$_3$OH/bath gas mixtures in the storage bulb. In the UR1 apparatus, between 14 and 21 kinetic experiments were performed for each Laval nozzle. 

\subsection{Chemical model}

In order to evaluate the role of reaction~(\ref{reac:1}) in the formation of CH$_3$O in B1-b, we have built a chemical model adopting typical conditions of a cold dense cloud, i.e., a temperature of 10 K, a visual extinction of 30 mag, a cosmic-ray ionization rate of H$_2$ of $1.3\times10^{-17}$ s$^{-1}$, and the so-called "low-metal" elemental abundances (see \citealt{agu2013}). According to \citet{dan2013}, the volume density of H$_2$ in B1-b has a steep radial gradient and takes values from a few $10^6$ cm$^{-3}$ at the core center to $\sim10^4$ cm$^{-3}$ in the outer regions of the cloud (at some tens of arcsec from the center). Since methanol has a widespread distribution in B1-b \citep{obe2010}, and methoxy has probably a similar distribution, we adopted densities of H nuclei of $2\times10^4$ cm$^{-3}$ and $1\times10^5$ cm$^{-3}$ to investigate the effect of density. Since in cold dense clouds, grain-surface chemistry is hampered by the low mobility of most species on ice surfaces at the low temperatures of dust, and since we aim at evaluating the efficiency of gas-phase routes to CH$_3$O, we did not consider chemical reactions on grain surfaces other than the formation of H$_2$ by recombination of two H atoms. We adopted the gas-phase reaction network of \citet{rua2015} with the rate coefficient for reaction~(\ref{reac:1}) updated according to this work. Since it is well known that pure gas-phase chemical models severely underestimate the gas-phase abundance of CH$_3$OH in cold dense clouds (e.g., \citealt{agu2013}), to be more realistic, we assumed an abundance of methanol relative to CO of $6\times10^{-5}$, which is the value in B1-b according to the abundances of CO and CH$_3$OH derived by \citet{lis2002} and \citet{obe2010}, respectively.

The kinetic information on the gas-phase reactivity of CH$_3$O is quite scarce and limited to temperatures above 200 K. Theoretical and experimental studies indicate that reactions of CH$_3$O with molecules such as H$_2$, CO, CH$_4$, NH$_3$, or CH$_3$OH have important activation barriers \citep{san1980,jod1999,wan1999}. The depletion of CH$_3$O in cold dense clouds is likely to be dominated by reactions with H and O atoms, 
\begin{subequations} \label{reac:6}
\begin{align}
\rm CH_3O + H & \rm \rightarrow H_2CO + H_2, \label{reac:6a} \\
                            & \rm \rightarrow CH_3 + OH, \label{reac:6b}
\end{align}
\end{subequations}
\begin{subequations} \label{reac:7}
\begin{align}
\rm CH_3O + O & \rm \rightarrow H_2CO + OH, \label{reac:7a} \\
                            & \rm \rightarrow CH_3 + O_2, \label{reac:7b}
\end{align}
\end{subequations}
whose rate coefficients are however not known at interstellar temperatures. We have adopted the same values used by \citet{rua2015}\footnote{Note that the rate coefficient of reaction~(\ref{reac:7b}) given in Table A2 of \citet{rua2015} should read $1.9\times10^{-11}$ cm$^3$ molecule$^{-1}$ s$^{-1}$ instead of $1.9\times10^{-12}$ cm$^3$ molecule$^{-1}$ s$^{-1}$.} for $k_{6a,b}$ and $k_{7a,b}$, which are based on measurements at 300 K \citep{hoy1981,dob1991,ewi1987,bau2005}. Therefore, in the absence of grain-surface processes, the chemistry of CH$_3$O becomes rather simple in the model as it is assumed to be formed through reaction~(\ref{reac:1}) and destroyed by reactions~(\ref{reac:6}) and (\ref{reac:7}). Therefore, at steady state the CH$_3$O/CH$_3$OH abundance ratio is given by:
\begin{equation}
\frac{\rm [CH_3O]}{\rm [CH_3OH]} = \frac{k_1 \rm [OH]}{k_6 \rm [H] + {\emph k_7} [O]}, \label{eq:8}
\end{equation}
where $k_1$ is well constrained by our experiments and thus the major source of uncertainty in the calculated CH$_3$O/CH$_3$OH abundance ratio comes from the values of $k_6$ and $k_7$ at low temperatures.

\section{Results and discussion}

\subsection{Evaluation of the role of secondary chemistry of methanol in the determination of $k_1(T)$}

The knowledge of the methanol concentration is one of the key parameters to derive a reliable rate coefficient from Equation~(\ref{eq:5}). For that reason, an evaluation of the potential effect of any side reaction that could deplete methanol is desirable.

No photolysis of methanol is expected at the photolysis (248 nm or 266 nm) and excitation (282 nm) wavelengths at the concentration levels and the low laser fluences (less than 1 mJ cm$^{-2}$ per laser pulse measured at the exit of the nozzle in UCLM and 75 mJ per laser pulse in UR1 measured at the exit of the laser) used in this work, if we consider an upper limit for the absorption cross section of 10$^{-22}$ cm$^2$ molecule$^{-1}$ \citep{che2002}.

As can be seen in Fig.~\ref{fig:1}, a slight curvature in the plots of $k'-k_0$ versus [CH$_3$OH] was observed at methanol concentrations higher than $1\times10^{14}$ molecule cm$^{-3}$ for the lowest temperature, 22 K. This curvature was also observed by \citet{gom2014} at 56 K for reaction~(\ref{reac:1}) for methanol concentrations higher than $2.5\times10^{14}$ molecule cm$^{-3}$. This may be explained by a clustering process that reduces the amount of methanol available for reacting with OH radicals, leading to a decrease of the measured pseudo-first order rate coefficient $k'$. The lower the temperature is, the more favourable the cluster formation is, shifting the curvature in the pseudo-first order plots to much lower methanol concentrations. \citet{lak2011} observed the formation of clusters of methanol, and even the formation of small droplets in the divergent section of a Laval nozzle at 210 K, although the methanol concentration in their experiments was around 4 orders of magnitude higher than in ours. Therefore, we believe that under our experimental conditions the clusters that are formed are mainly methanol dimers. In order to estimate the effect of dimerization in the measurement of the pseudo-first order coefficients, a simple model has been developed using FACSIMILE software for the worst scenario, i.e., the lowest temperature where the gas density is the highest (see Table~\ref{tab:2}). The reaction scheme of this model includes reactions~(\ref{reac:1}) and (\ref{reac:3}) together with the formation of a hydrogen-bonded methanol dimer:
\begin{equation}
\rm 2~CH_3OH \rightarrow (CH_3OH)_2. \label{reac:9}
\end{equation}

\begin{deluxetable}{ccc}
\tablecaption{Gas-phase rate coefficients of the reaction between OH and CH$_3$OH determined in this work \label{tab:2}} \tablewidth{0pc}
\startdata \hline \hline
\multicolumn{1}{c}{$T$} & \multicolumn{1}{c}{$n$} & \multicolumn{1}{c}{$k_1$} \\
\multicolumn{1}{c}{(K)} & \multicolumn{1}{c}{(10$^{16}$ cm$^{-3}$)} & \multicolumn{1}{c}{(10$^{-11}$ cm$^3$ molecule$^{-1}$ s$^{-1}$)} \\
\hline
22.4$\pm$1.4  & 17.0$\pm$1.6    & 4.30$\pm$0.66 \\
42.5$\pm$1.3  & 5.22$\pm$0.33  & 2.74$\pm$0.42 \\
47.7$\pm$0.6  & 2.74$\pm$0.09  & 2.17$\pm$0.35 \\
51.6$\pm$1.7  & 4.17$\pm$0.35  & 2.19$\pm$0.26 \\
52.2$\pm$0.9  & 5.15$\pm$0.13  & 2.97$\pm$0.60 \\
61.0$\pm$1.0  & 2.02$\pm$0.08  & 2.20$\pm$0.40 \\
64.2$\pm$1.7  & 2.24$\pm$0.15  & 1.47$\pm$0.23
\enddata
\tablecomments{Uncertainties are the combination of statistical error ($\pm t~\sigma$, Student $t$ factor at $95\%$ confidence level) and systematic errors ($15\%$).}
\end{deluxetable}

In the simulation, $k_1(T)$ and the pseudo-first order $k'$ in the absence of methanol were fixed to the value obtained from the linear fit according to Equation~(\ref{eq:5}) and the experimental $k_0$. It was also considered that the dimer does not react with OH radicals. We estimate that the observed curvature is compatible with a rate coefficient for the dimerization of methanol at 22 K of the order of $10^{-11}$ cm$^3$ molecule$^{-1}$ s$^{-1}$. The effect of dimerization on $k'$ can be neglected at concentrations lower than $1\times10^{14}$ molecule cm$^{-3}$, the ones considered in this study. For intermediate temperatures, the concentration range was also constrained to [CH$_3$OH] below $2\times10^{14}$ molecule cm$^{-3}$, while at 64 K the maximum methanol concentration was kept below $3.5\times10^{14}$ molecule cm$^{-3}$. Consequently, only the linear part of these plots were considered in obtaining $k_1(T)$.

\subsection{Potential role of water in the determination of $k_1(T)$}

Water, which was introduced in conjunction with H$_2$O$_2$ in the reactor, did not play any significant role in the chemistry of the flow. Its potential aggregation to methanol to form a H$_2$O$\cdots$CH$_3$OH complex was found to be a negligible process in the loss of methanol. Indeed, water concentration in the cold flow was estimated to be similar to that of H$_2$O$_2$ taking into account the concentration ratios in the liquid mixture and the vapor pressures of H$_2$O and H$_2$O$_2$, 42.4 mbar and 24.4 mbar at 300 K, respectively. The H$_2$O$_2$ concentration was evaluated using a room temperature UV absorption column set in between the H$_2$O$_2$ bubbler and the reservoir at the pressure condition of our experiment at 64 K. The obtained concentration converted into the supersonic flow conditions was found to be about $6\times10^{12}$ cm$^{-3}$, representing a few percent of the typical [CH$_3$OH] used in this experiment (see Fig.~\ref{fig:1}). The work of \citet{voh2007} on the catalytic effect of water on the OH+acetaldehyde gas-phase reaction between 58 K and 300 K indicates that at temperatures below 100 K the complexation of acetaldehyde molecules with water is highly efficient, although at high water concentrations ($3~\%$ of the total density, i.e., $2.55\times10^{15}$ cm$^{-3}$ at 77 K). In the work of \citet{voh2007}, the [H$_2$O]/[CH$_3$CHO] ratio is around 16. In our work at 64 K, [H$_2$O] in the supersonic jet is estimated to be on the order of $10^{13}$ cm$^{-3}$, and therefore [H$_2$O]/[CH$_3$OH] ratio is $<0.1$. Water content accounts for 0.04~\% of the total gas density ($2.24\times10^{16}$ cm$^{-3}$), i.e., it is negligible. Moreover, \citet{voh2007} concluded that at water contents lower than 3~\% no catalytic effect was observed for the OH+acetaldehyde system. Under our experimental conditions, if H$_2$O$\cdots$CH$_3$OH cluster is formed with an aggregation factor of unity (as suggested by \citealt{voh2007} at $T<100$~K) and assuming that the kinetics of the dimer formation is fast enough to produce it in the timescale of our experiment (300 microseconds), the methanol concentration would only be slightly reduced, thus not affecting the measured rate coefficient.

\subsection{Rate coefficients $k_1(T)$ between 22 K and 64 K}

\begin{figure}
\centering
\includegraphics[angle=0,width=\columnwidth]{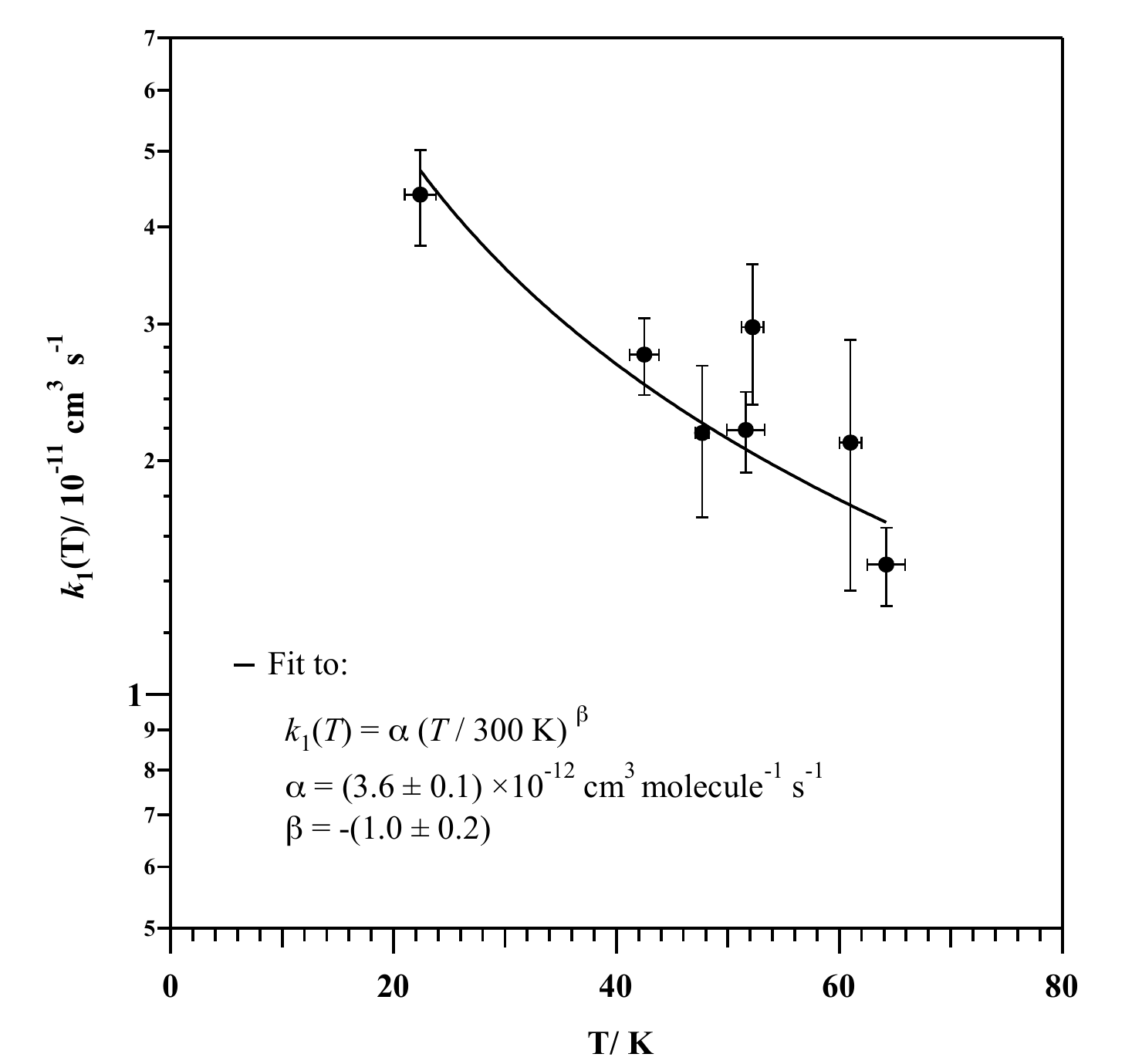} \caption{Rate coefficients for the reaction of OH with CH$_3$OH as a function of temperature determined in this work between 22 and 64 K. The black line corresponds to Equation~(\ref{eq:10}).} \label{fig:2}
\end{figure}

In light of the absence of any secondary chemistry under our experimental conditions, $k_1(T)$ were obtained, as explained in Section~\ref{sec:methods}, and are shown in Table~\ref{tab:2}. Statistical errors in $k_1(T)$ result from a least squares analysis of the pseudo first order rate coefficient multiplied by the Student-$t$ factor corresponding to the $95\%$ confidence limit. The errors given in Table~\ref{tab:2} include as well a contribution of systematic errors, estimated to be about $15\%$, which essentially come from uncertainties in mass flows or pressure gauge calibrations and fluctuations in the temperature and concentration of methanol. The total gas densities are also listed in Table~\ref{tab:2}. Note that the study of any pressure dependence of the rate coefficient is not possible for a single Laval nozzle operating with a specific bath gas and at a fixed temperature. Nevertheless, no pressure dependence of $k_1(T)$ seems to exist at ultra-low temperatures, according to \citet{sha2013}, so the results obtained for the temperatures employed in this work at different gas densities can be compared.

The rate coefficients $k_1(T)$ obtained in this work between 22 and 64 K are depicted in Fig.~\ref{fig:2} in a $\log-\log$ form. It is seen that $k_1(T)$ increases as the temperature decreases, i.e., it shows a negative temperature dependence. The data points have been fitted in the temperature range 22-64 K to the expression:
\begin{equation}
k_1(T) = (3.6\pm0.1)\times10^{-12} \Big(\frac{T}{300~{\rm K}}\Big)^{-(1.0\pm0.2)}, \label{eq:10}
\end{equation}
where $k_1(T)$ has units of cm$^3$ molecule$^{-1}$ s$^{-1}$. The extrapolation of Equation~(\ref{eq:10}) to a temperature of 10 K, of interest for cold dark clouds such as B1-b, yields a value of $1.1\times10^{-10}$ cm$^3$ molecule$^{-1}$ s$^{-1}$. It is worth noting that for the present limited experimental temperature range, several mathematical expressions give fairly acceptable fits. However, for some of these expressions the extrapolation beyond the experimental temperature range, especially at temperatures significantly below 22 K, can diverge or result in unrealistic rate coefficients. Future works, which require the construction of new Laval nozzles, are planned to extend the present temperature range in order to better constrain the dependence with temperature of $k_1$.

\begin{figure}
\centering
\includegraphics[angle=0,width=\columnwidth]{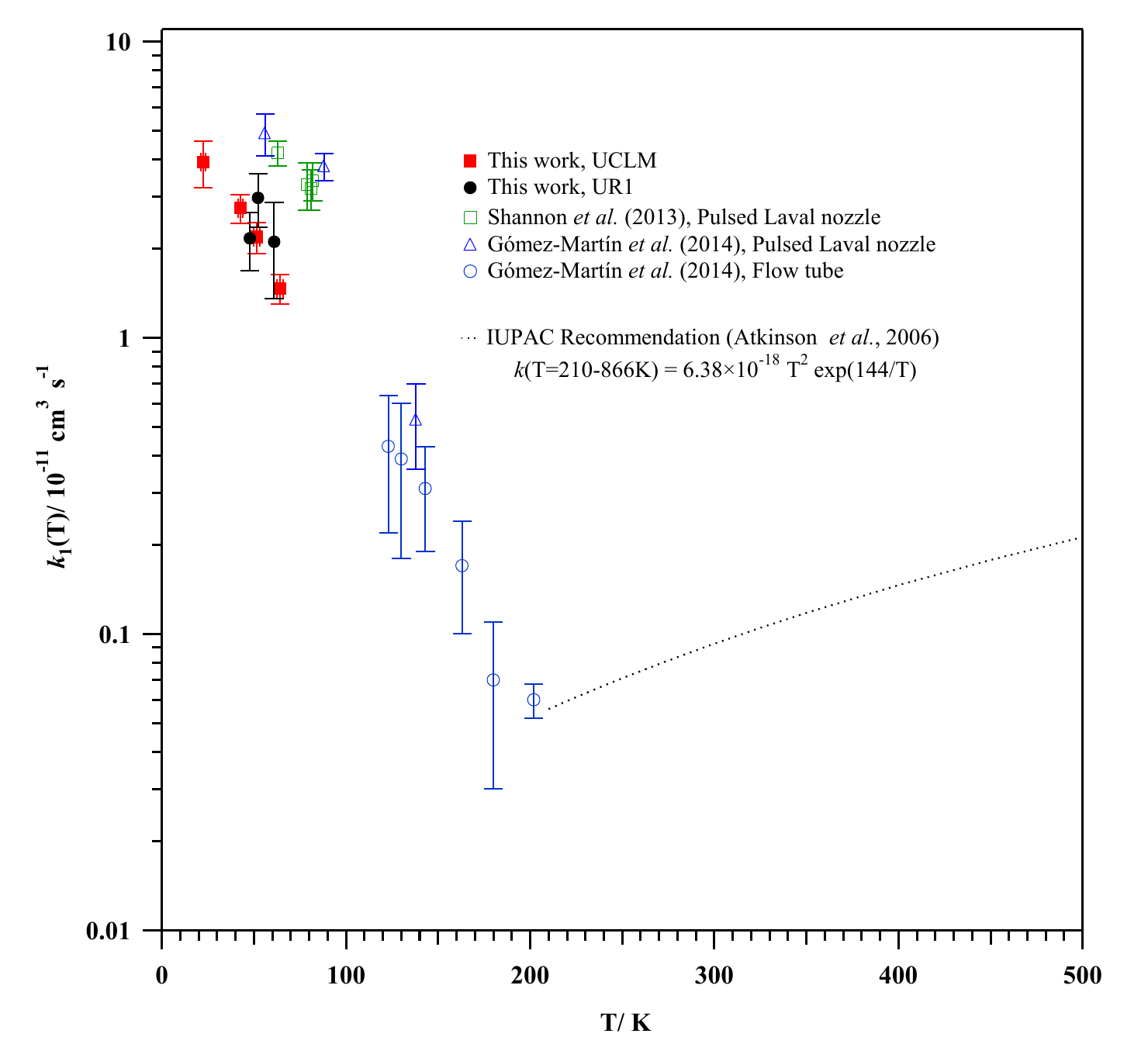} \caption{Rate coefficients for the reaction of OH with CH$_3$OH as a function of temperature. Filled symbols are kinetic data presented in this work and empty symbols are values reported in the literature.} \label{fig:3}
\end{figure}

For comparison purposes, values of  $k_1$ reported at temperatures higher than 56 K are depicted in Fig.~\ref{fig:3}. The methanol reactivity toward OH radicals at 22 K is enhanced by almost two orders of magnitude compared to 200 K \citep{gom2014}. The negative temperature dependence of $k_1(T)$ was already observed by \citet{sha2013} and \citet{gom2014} in the 56-202 K range. Nevertheless, the values of $k_1(T)$ previously determined by these authors using a pulsed Laval nozzle are around twice higher than ours in the temperature range common to these authors and us (close to 60 K). The source of this discrepancy is presently unknown and should deserve more experimental attention. The good agreement observed in the present study employing two independent CRESU machines shed confidence to the values reported here. Furthermore, the trend observed in this work between 22 and 64 K is consistent with the results obtained by \citet{gom2014} between 123 and 202 K using a classical flow tube coupled to a PLP-LIF technique.

At temperatures between 210 and 866 K the trend in the temperature dependence of $k_1(T)$ is inverted, showing a positive temperature dependence \citep{atk2006}. This opposite behaviour yields a "U-shape" in the $\log k_1(T)$ versus $T$ plots, already observed for other reactions of OH with complex organic molecules \citep{jim2015a,jim2015b,sha2010,sha2014,car2015}. At ultra-low temperatures, the reaction mechanism that allows this to occur is one in which an adduct is formed by a weak hydrogen-bonded association of OH and methanol followed by quantum-mechanical tunnelling, as suggested by \citet{sha2013}. Recently, \citet{her2015} reported experimental evidence of the formation of the hydrogen bonded intermediate OH$\cdots$CH$_3$OH by He Nanodroplet Isolation (HENDI) and a combination of mass spectrometry and infrared laser Stark spectroscopy.

\subsection{Predicted gas-phase abundance of CH$_3$O}

As stated in Section~\ref{sec:intro}, the reaction of OH with CH$_3$OH may have important implications for interstellar chemistry since it acts as a sink of methanol and a source of CH$_3$O radicals. In particular, here we are interested in evaluating whether this reaction can explain the observed abundance of CH$_3$O in the cold dense core B1-b, $\sim5\times10^{-12}$ relative to H$_2$ or $\sim3\times10^{-3}$ with respect to CH$_3$OH \citep{cer2012}.

\begin{figure}
\centering
\includegraphics[angle=0,width=\columnwidth]{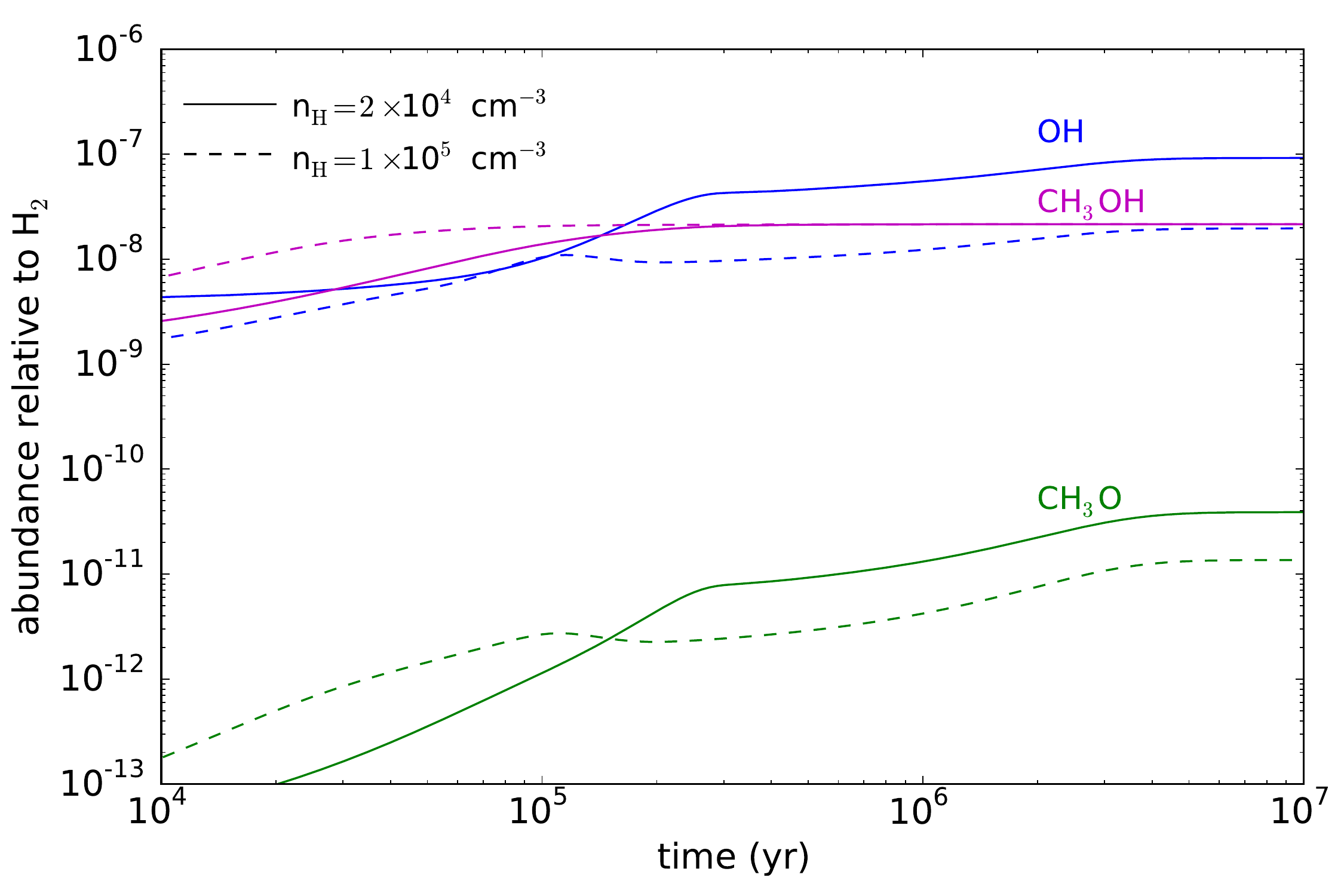} \caption{Calculated abundances of OH, methanol and CH$_3$O relative to H$_2$ are shown as a function of time. Solid and dashed lines correspond to densities of H nuclei of $2\times10^4$ cm$^{-3}$ and $1\times10^5$ cm$^{-3}$, respectively.} \label{fig:4}
\end{figure}

Figure~\ref{fig:4} shows the time evolution of the predicted fractional abundances of the reactants involved in reaction~(\ref{reac:1}), OH and CH$_3$OH, and the reaction product CH$_3$O for two different densities of H nuclei. It is seen that methanol reaches its maximum abundance soon after a few $10^4$ yr (remember that the abundance of CH$_3$OH has been fixed to $6\times10^{-5}$ relative to CO), while the methoxy radical increases its abundance progressively until reaching the maximum at late times. The range of abundances of CH$_3$O relative to methanol, $n$(CH$_3$O)/$n$(CH$_3$OH), achieved within a time range between $10^5$ yr and $10^7$ yr are listed in Table~\ref{tab:3}. Our chemical model predicts that at $\sim10^7$ yr, $n$(CH$_3$O)/$n$(CH$_3$OH) reaches a value of $2\times10^{-3}$ or $0.6\times10^{-3}$, depending on the adopted density of H nuclei. This ratio is somewhat lower, although of the same order, than the one observed in B1-b. The main uncertainty in the calculated  CH$_3$O/CH$_3$OH abundance ratio arises from the depletion rate of CH$_3$O, which in turn depends on the rate coefficients of the reactions of CH$_3$O with H and O atoms and on the abundances of H and O atoms, both species being very difficult to observe in cold dense clouds. Nonetheless, with current constraints on the destruction rate of CH$_3$O, the title reaction is rapid enough at low temperatures to provide an efficient pure gas-phase pathway to CH$_3$O radicals in B1-b. Thus, our results indicate that gas-phase formation of CH$_3$O via reaction~(\ref{reac:1}), even though it does not fully interpret the observed abundance of methoxy radicals in B-1b, is likely to be a major contributor. Up to now, it is thought that the main competing process of formation of CH$_3$O is its synthesis on grain surfaces and further non-thermal desorption to the gas phase. As \citet{cer2012} pointed out, experiments that simulate ice mantle processing by ultraviolet or ion irradiation indicate that it is not the CH$_3$O isomer, but the more stable isomer CH$_2$OH, which is formed on grain surfaces. This was proven by \citet{che2013}, who observed CH$_2$OH in the X-ray irradiation of methanol ice at 14 K. The warming up of these ices produces ethylene glycol (CH$_2$OH)$_2$ indicating that the abundance of CH$_2$OH radicals is high. More recently, \citet{lee2015} observed that CH$_3$O formed in the 355 nm irradiation of a $p$-H$_2$ matrix containing methyl nitrite (CH$_3$ONO) at 3.2 K is rapidly converted into CH$_2$OH. These authors theoretically predict that the conversion of CH$_3$O to CH$_2$OH on grain surfaces may occur in two steps:
\begin{equation}
\rm CH_3O + H_2 \rightarrow CH_3OH + H, \label{reac:11}
\end{equation}
\begin{equation}
\rm CH_3OH + H \rightarrow CH_2OH + H_2. \label{reac:12}
\end{equation}
In light of these results, the formation of CH$_3$O through grain-surface processes seems to be disfavoured, although more laboratory experiments and/or quantum chemistry calculations are needed to draw definitive conclusions.

\subsubsection{Comparison with previous chemical models}

\begin{deluxetable*}{lcrrcl}
\tablecaption{Calculated abundances of CH$_3$O relative to methanol at 10 K considering gas-phase formation via reaction~(\ref{reac:1}) \label{tab:3}} \tablewidth{0pc}
\startdata \hline \hline
\multicolumn{1}{c}{Grain surface} & \multicolumn{1}{c}{$n_{\rm H}$} & \multicolumn{1}{c}{$k_1$(10 K)} & \multicolumn{1}{c}{$n$(CH$_3$O)/$n$(CH$_3$OH)} & \multicolumn{1}{c}{time} & \multicolumn{1}{l}{Reference} \\
\multicolumn{1}{c}{source of CH$_3$O} & \multicolumn{1}{c}{(molecule cm$^{-3}$)} & \multicolumn{1}{c}{(cm$^3$ molecule$^{-1}$ s$^{-1}$)} & \multicolumn{1}{c}{} & \multicolumn{1}{c}{(yr)} & \multicolumn{1}{c}{} \\
\hline
NO   & $2\times10^4$  & $1.1\times10^{-10}$  & $(0.08-2)\times10^{-3}$    & $10^5-10^7$ & This work \\
NO   & $1\times10^5$  & $1.1\times10^{-10}$  & $(0.1-0.6)\times10^{-3}$   & $10^5-10^7$ & This work \\
YES & $2\times10^4$  & $3\times10^{-10}$     & $\sim5000\times10^{-3}$  & $10^5-10^7$ & \citet{ach2015} \\
YES & $1\times10^5$  & $4\times10^{-11}$     & $(10-100)\times10^{-3}$   & $10^5-10^7$ & \citet{rua2015} \\
NO   & $6\times10^4$  & $3\times10^{-10}$     & $\sim40\times10^{-3}$      & $\sim10^5$   & \citet{bal2015} \\
NO   & $1\times10^5$  & $4\times10^{-11}$     & $(2-10)\times10^{-3}$       & $10^5-10^6$ & \citet{vas2013} \\
YES & $1\times10^5$  & $4\times10^{-11}$     & $(0.3-50)\times10^{-3}$    & $10^5-10^6$ & \citet{vas2013}
\enddata
\tablecomments{The observed $n$(CH$_3$O)/$n$(CH$_3$OH) in B1-b is $\sim3\times10^{-3}$ \citep{obe2010,cer2012}.}
\end{deluxetable*}

In Table~\ref{tab:3}, some input parameters used in previous chemical models and the predicted abundances of CH$_3$O are listed for comparison purposes. The first column of the table indicates whether or not the model includes grain surface production of CH$_3$O radicals and the second column provides the density of H nuclei adopted in each model. As can be seen in the third column, the rate coefficient $k_1$(10 K) used in previous models, either $4\times10^{-11}$ cm$^3$ molecule$^{-1}$ s$^{-1}$ (measured by \citealt{sha2013} at 63 K) or $3\times10^{-10}$ cm$^3$ molecule$^{-1}$ s$^{-1}$ (collision limit), differs by almost one order of magnitude. The fourth column gives the range of CH$_3$O/CH$_3$OH abundance ratios predicted by each model over a certain time interval, indicated in the fifth column.

Our results are consistent with those of \citet{vas2013} when grain-surface production of CH$_3$O is switched off in their model and reaction~(\ref{reac:1}) becomes the only efficient formation route to CH$_3$O. The CH$_3$O/CH$_3$OH abundance ratios reported by \citet{vas2013} are around or slightly above $10^{-3}$, while ours are around or slightly below $10^{-3}$. \citet{bal2015} do only consider gas-phase formation of CH$_3$O with $k_1$(10 K) = $3\times10^{-10}$ cm$^3$ molecule$^{-1}$ s$^{-1}$ and find a CH$_3$O/CH$_3$OH abundance ratio of $\sim4\times10^{-2}$, about 10 times higher than the value observed in B1-b, over a restricted time interval around $10^5$ yr.

As seen in Table~\ref{tab:3}, a common outcome of models that consider formation of CH$_3$O on grain surfaces \citep{vas2013,rua2015,ach2015} is a significant overestimation of the CH$_3$O/CH$_3$OH abundance ratio with respect to the observed value in B1-b. This fact points to an overestimation in the formation efficiency of CH$_3$O through grain-surface routes, which is consistent with the empirical conclusion that CH$_3$O is not predominantly formed on grains surfaces. It must be noted that the efficiency of the grain-surface processes invoked to explain the formation of CH$_3$O is highly uncertain.

In the chemical models of \citet{vas2013} and \citet{rua2015}, CH$_3$O is assumed to be formed mainly on the surface of dust grains through successive hydrogenations of CO. Methoxy radicals are assumed to be further ejected to the gas phase through chemical desorption, which assumes that a fraction, typically $1\%$, of the energy released by exothermic chemical reactions on grain surfaces is used by the products to desorb. In the chemical model of \citet{rua2015}, CH$_3$O is also formed through a mechanism involving the formation of van der Waals complexes between carbon atoms colliding with dust grains and water ice molecules. In these two chemical models, the formation of CH$_3$O through the gas-phase reaction of OH and CH$_3$OH was included although grain-surface processes are found to dominate the formation of CH$_3$O. The gas-phase production rate of CH$_3$O is similar in both models, although the depletion of CH$_3$O through reactions with H and O atoms is assumed to be faster in the model of \citet{vas2013}. This makes that, as seen in Table~\ref{tab:3}, the values of $n$(CH$_3$O)/$n$(CH$_3$OH) predicted by \citet{rua2015} are higher than those calculated by \citet{vas2013}.

\citet{ach2015} modelled the fractional abundance of CH$_3$O using an updated gas-grain 
chemical model that included the gas-phase reaction of OH with methanol (reaction~\ref{reac:1}) and with some other organic molecules. These authors run the model with and without reaction~(\ref{reac:1}) and concluded that at 10 K this reaction is not generally an effective destruction route for CH$_3$OH, although it leads to an increase in the abundance of CH$_3$O at times longer than $2\times10^5$ yr. The main competing process of formation of gas-phase CH$_3$O in this model is the hydrogenation of formaldehyde on grain surfaces followed by non-thermal desorption. Clearly, in this model the CH$_3$O/CH$_3$OH abundance ratio is largely overestimated with respect to the observed value in B1-b, probably as a consequence of a too efficient formation of CH$_3$O via grain-surface routes.

\section{Conclusions}

In this work, we have determined the rate coefficient $k_1$ for the gas-phase reaction between OH radicals and CH$_3$OH in the range of temperatures between 22 and 64 K. This study provides the first measurement of $k_1$ at temperatures below 56 K. The lowest temperature investigated here, 22 K, is the closest one to that in B1-b at which the rate coefficient of the reaction between OH and CH$_3$OH has been measured. In the cold environment of B1-b, CH$_3$O (product of the reaction between OH and CH$_3$OH) has been detected recently. The consistency of the measured rate coefficients in two different laboratories (UCLM and UR1) leads us confidence in the reported $k_1(T)$, especially around 60 K where there is some discrepancy with previous data. We fit the measured rate coefficients as a function of temperature and provide an expression of $k_1(T)$ in the temperature range 22-64 K. Further measurements of $k_1$ at temperatures below 22 K are needed to corroborate whether this reaction continues to be accelerated at even lower temperatures and to constrain by how much in order to better evaluate its contribution to the production of CH$_3$O in cold dense clouds such as B1-b. In addition, there is still a gap to fill between 64 and 123 K, which is planned to be covered in the near future with new Laval nozzles that will help to elucidate the kinetic behaviour at intermediate temperatures. Finally, we will intend to detect CH$_3$O radicals formed in the reaction of OH with methanol at 22 K by LIF. This is planned to be done in the near future, since some changes are needed in the experimental system to detect CH$_3$O.

We have also modeled the chemistry of CH$_3$O in cold dense clouds considering that it is only formed through the gas-phase reaction of OH and CH$_3$OH. We have adopted the extrapolated value of $k_1(T)$ at 10 K and assumed that gas-phase reactions with H and O atoms are the only removal processes of CH$_3$O. This model predicts CH$_3$O/CH$_3$OH abundance ratios that are similar or slightly below the value observed in B1-b ($\sim3\times10^{-3}$), confirming that the reaction of OH and CH$_3$OH is clearly involved in the formation of interstellar methoxy radicals. This is consistent with recent experiments that indicate that CH$_2$OH, rather than CH$_3$O, is the isomer formed on icy grain mantles. Nonetheless, the role of grain-surface processes on the chemistry of CH$_3$O needs to be further investigated. Probably a search for the more stable isomer, CH$_2$OH, in cold dense clouds may shed light on the relative role of gas-phase and grain-surface chemical reactions in the synthesis of methanol-derived radicals. Up to date, the rotational spectrum of CH$_2$OH has not been yet accurately measured in the laboratory, which complicates the detection in the interstellar medium.

\acknowledgements

This work has been supported by the European Research Council and the Spanish Ministry of Science and Innovation through the NANOCOSMOS (SyG-610256) and ASTROMOL (CSD2009-00038) projects, respectively. Authors from UCLM and CSIC acknowledge the Spanish Ministry of Economy and Competitiveness for supporting this work under projects GASSOL (CGL2013-43227-R), AYA2009-07304, and AYA2012-32032. Authors from UR1 thank the French national programme PCMI "Physique et Chimie du Milieu Interstellaire" from INSU (Institut National des Sciences de l'Univers) for financial support. A. Canosa is grateful to UCLM for providing him with an invited researcher position for 5 months. Authors thank G. M. Mu\~noz Caro for helpful discussions.

\end{document}